\documentclass[aps,prd,twocolumn,superscriptaddress]{revtex4}

\usepackage{graphicx}
\usepackage{dcolumn}
\usepackage{doi}
\usepackage{amsmath}
\usepackage{amssymb}
\usepackage{bm}
\usepackage{color}
\usepackage[normalem]{ulem}
\usepackage[dvipsnames]{xcolor}
\usepackage{hyperref}
\hypersetup{
colorlinks=true,
linkcolor=blue,
citecolor=magenta,
}

\begin{document}
\title{Quasinormal modes of magnetized black hole}
\author{Bobur Turimov}
\email{bturimov@astrin.uz} \affiliation{Ulugh Beg Astronomical
Institute, Astronomicheskaya 33,Tashkent 100052, Uzbekistan}
\affiliation{Institute of Physics and Research Centre of
Theoretical Physics and Astrophysics, Faculty of Philosophy and
Science, Silesian University in Opava, Bezrucovo namesti 13,
CZ-74601 Opava, Czech Republic}
\author{Bobir Toshmatov}
\email{toshmatov@astrin.uz} \affiliation{Department of Physics,
Nazarbayev University,  53 Kabanbay Batyr, 010000 Nur-Sultan,
Kazakhstan} \affiliation{Ulugh Beg Astronomical Institute,
Astronomicheskaya 33,Tashkent 100052, Uzbekistan}
\affiliation{Institute of Physics and Research Centre of
Theoretical Physics and Astrophysics, Faculty of Philosophy and
Science, Silesian University in Opava, Bezrucovo namesti 13,
CZ-74601 Opava, Czech Republic}
\author{Bobomurat Ahmedov}
\email{ahmedov@astrin.uz} \affiliation{Ulugh Beg Astronomical
Institute,  Astronomicheskaya 33,Tashkent 100052, Uzbekistan}
\affiliation{National University of Uzbekistan,Tashkent 100174,
Uzbekistan}
\author{Zden{\v e}k Stuchl{\'i}k}
\email{zdenek.stuchlik@physics.cz} \affiliation{Institute of
Physics and Research Centre of  Theoretical Physics and
Astrophysics, Faculty of Philosophy and Science, Silesian
University in Opava, Bezrucovo namesti 13, CZ-74601 Opava, Czech
Republic}
\date{\today}
\begin{abstract}
We investigate charged, massive scalar field around static,
spherically symmetric black hole immersed into an external
asymptotically uniform magnetic field $B$. It is shown that for
given multipole number $\ell$ there are $2\ell+1$ numbers of modes
due to the Zeeman effect appearing by an interaction of the
external magnetic and charged scalar fields introducing an
effective mass of the scalar field $\mu_{\rm
eff}=\sqrt{\mu^2-mqB}$ where $m$ is the azimuthal number and $q$
is the charge coupling constant. We calculate threshold value of
effective mass in which quasinormal modes are arbitrarily long
lived and beyond that value quasinormal modes vanish. In the case
of $m qB<0$ quasinormal modes are longer lived with larger
oscillation frequencies. Whenever, magnetic and massive scalar
fields satisfies condition $\mu_{\rm eff}^2<0$, an instability
appears, i.e., if $qB>0$ or $qB<0$ there is an instability for the
values of azimuthal number $m>\mu^2/qB$ or $m<\mu^2/qB$,
respectively.
\end{abstract}


\maketitle

\section{Introduction}

According to ``no-hair theorem" black hole does not possess its
own magnetic field. In the pioneering papers
\cite{Giznburg:Ozernoy,Ginzburg:1964} Ginzburg and Ozernoy have
shown that during the gravitational collapse the dipolar magnetic
moment of the star will decay vice proportional with time and
estimation has been made more accurate for multipolar magnetic
fields by Price~\cite{Price:PRD:1972}. However, the external
magnetic field around the black hole can be generated by an
accretion disc, or a rotating ring of matter that falls into the
black hole, or a companion in binary systems containing neutron
star or magnetar with strong magnetic field. The black hole
immersed in an asymptotically uniform external magnetic field has
been first considered by Wald~\cite{Wald74} and several physical
scenarios have been considered later studied, e.g.,
in~\cite{Kolos:CQG:2015,Toshmatov:APSS:2015,Stuchlik:EPJC:2016,Tursunov:PRD:2016,TursunovAPJ,Morozova:PRD,Rayimbayev:APSS,Kolos:EPJC:2017,Frolov:PRD:2013,AAA:PS:2014,AAA:PRD:2013,Turimov18a,Shaymatov:IJMPD:2018,AAA:PRD:2019}.
It is generally accepted that a magnetic field is considered one
of main sources of the most energetic processes around
supermassive black holes at the center of galaxies, playing the
role of ``feeder" of the supermassive black hole by trapping dust
near the galaxy’s center~\cite{Sofia:2018}.

A complete detailed analysis of interaction between black hole and
magnetic field generated by the accretion disc or companion object
(it can be neutron star or magnetar with strong magnetic field) is
complicated problem which requires numerical magnetohydrodynamic
(MHD) simulations~\cite{McKinney:MNRAS}. However, approximative
methods are also very useful to draw picture of this phenomenon,
by considering stationary magnetized black hole solutions in
general relativity as was done by Wald~\cite{Wald74} and
Ernst~\cite{Ernst:JMP:1976}.

The study of an interaction between scalar and electromagnetic
fields is very interesting topic from theoretical and
observational point of view. A comprehensive physical aspects of
the theory of black holes in an external electromagnetic field are
reviewed in~\cite{Aliev:1989,Kokkotas:PRD:2011}. In the
papers~\cite{Konoplya08a,Konoplya08b} propagation of scalar field
in the background of strongly magnetized black hole (or Ernst
spacetime) has been studied and later it is considered for the
massive scalar field in~\cite{Wu15} in $G^{1/2}c^{-2}Br\ll1$. It
is shown that in the presence of the strong magnetic field the
quasinormal modes are longer lived and have larger oscillation
frequencies in both massless and massive scalar
fields~\cite{Brito14}. However, such approximation is not accurate
enough in the strong magnetic and gravitational fields when
$G^{1/2}c^{-2}Br\sim1$. Moreover, it was shown in
\cite{Frolov:PRD:2010} that the spacetime local curvature created
by the magnetic field $B$ can be neglected and it can be
considered as test magnetic field if strength of the magnetic
field satisfies the condition
\begin{eqnarray}\label{b-condition}
B\ll B_M=\frac{c^4}{G^{3/2}M_\odot} \left(\frac{M_\odot}{M}\right)
\sim10^{19} \left(\frac{M_\odot}{M}\right) {\rm G}\ .
\end{eqnarray}
On the other hand, observations have shown that the surface
magnetic  field of magnetar that is considered the most magnetic
object known in the Universe is in the order of $B\simeq
10^{15}{\rm G}$~\cite{Thompson:Apj:1992,Thompson:Apj:1993} which
is very small to change the geometry. Therefore, the Ernst
solution of the field equations has more academic interest than
the astrophysical relevance.

In the present paper we study massive and charged scalar field  in
the spacetime of Schwarzschild black hole immersed into the
uniform magnetic field assuming that the magnetic field satisfies
the condition (\ref{b-condition}). The paper is organized as
follows. In section \ref{sec:KG} we briefly describe Klein-Gordon
equation for the massive and charged field in the presence of the
electromagnetic field in the spacetime of static, spherically
symmetric black hole. In section \ref{Result} we present
quasinormal modes calculated by using the
Wentzel-Kramers-Brillouin (WKB) and Leaver's methods. Finally, we
present some concluding remarks in section \ref{Conclusion}.
Throughout the paper we use the geometric system of units $c = G =
\hbar = 1$ and spacelike signature ($-, +, +, +$).

\section{Basic equations\label{sec:KG}}

In this section we present equations of motion for the  massive
charged scalar field around the black hole immersed in the uniform
magnetic field and examine a contribution of the external magnetic
field on the quasinormal modes of the test scalar field. In the
spherical coordinates ($t,r,\theta,\phi$), line element of the
Schwarzschild black hole is written as
\begin{eqnarray}\label{metric}
ds^2=-f(r)dt^2 +\frac{dr^2}{f(r)}+r^2d\theta^2
+r^2\sin^2\theta d\phi^2\ ,
\end{eqnarray}
with metric function
\begin{eqnarray}
f(r) = 1-\frac{2M}{r}\ ,
\end{eqnarray}
where $M$ is total mass of the black hole.

In the paper~\cite{Wald74} the electromagnetic field configuration
in the vicinity of black hole that is immersed in the external
asymptotically uniform magnetic field has been studied. By
following that method we find the vector potential $A_\mu$ of the
electromagnetic field as
\begin{equation}\label{A}
A_\mu = \frac{1}{2}Br^2\sin ^2\theta \,(0,0,0,1)  \ ,
\end{equation}
where $B$ is the strength of uniform magnetic field.

The general relativistic form of Klein-Gordon equation for  the
massive and charged scalar field $\Phi$ in presence of the
electromagnetic field is given by
\begin{equation}\label{KG}
g^{\alpha\beta}\left(\nabla_\alpha - i q\, A_\alpha \right)
\left(\nabla_\beta - iq\, A_\beta \right)\Phi -\mu^{2} \Phi = 0 \ ,
\end{equation}
where $\mu$ is mass of the scalar field, $q$ is the charge
coupling constant between the scalar and electromagnetic fields,
$\nabla_\alpha$ is the covariant derivative.

Obviously, it is very difficult to separate variables in  equation
(\ref{KG}), however we can use the following physically reasonable
assumptions which facilitate the problem:
\begin{itemize}
\item Lorentz gauge for the vector potential $\nabla_\alpha A^\alpha = 0$;
\item In weak interaction limit one can ignore high order terms starting $q^2B^2 \to 0$.
\end{itemize}
Then, equation (\ref{KG}) can be expressed in the following simple form:
\begin{equation}\label{KGexpand}
\frac{1}{r^2}\frac{\partial}{\partial r}
\left(r^2f\frac{\partial\Phi}{\partial r}\right)
-\left(\frac{{\bf L}^2}{r^2}+\mu^{2}-qB\,L_z\right) \Phi
-\frac{1}{f}\frac{\partial^2\Phi}{\partial t^2} = 0\ ,
\end{equation}
where ${\bf L}^2$, $L_z$ are square of the total orbital
angular momentum and $z$ component of the orbital angular
momentum operators that have the following forms:
\begin{eqnarray}
{\bf L}^2 &=& -\left[\frac{1}{\sin\theta}
\frac{\partial}{\partial\theta}
\left(\sin\theta\frac{\partial}{\partial\theta}\right)
+\frac{1}{\sin^2\theta}\frac{\partial^2}{\partial\phi^2} \right] \ ,
\\
L_z &=& - i \frac{\partial}{\partial\phi} \ .
\end{eqnarray}
One can see that equation (\ref{KGexpand}) can be separated  into
radial and angular variables parts if the wave function is chosen
as harmonically time dependent form as
\begin{eqnarray}\label{Separation}
\Phi(t,r,\theta,\phi) = \frac{1}{2\pi}\int d\omega
e^{-i\omega t} \sum_{\ell}\frac{R_{\ell m}(r,\omega)}{r}
Y_{\ell m}(\theta,\phi) \ ,
\end{eqnarray}
where $\omega$ is the frequency of quasinormal modes of the
massive charged scalar field, $R_{\ell m}(r)$ is radial part of
the wave function. The angular part of the wave function $Y_{\ell
m}(\theta,\phi)$ is the spherical harmonic function which
satisfies the following relations:
\begin{eqnarray}
&&{\bf L}^2\,Y_{\ell m}(\theta,\phi) = \ell(\ell+1)\,Y_{\ell m}(\theta,\phi)\ ,
\\
&&L_z \,Y_{\ell m}(\theta,\phi) = m \,Y_{\ell m}(\theta,\phi) \ ,
\end{eqnarray}
where $\ell$ is the multipole number $\ell=0,1,2,...$ and $m$ is
the  azimuthal number with range $m\leq |\ell|$. Substituting
expression (\ref{Separation}) into (\ref{KGexpand}) one can write
the following stationary Schr\"{o}dinger-like wave equation for
the radial function:
\begin{equation}\label{WaveEq}
\left(\frac{d^2}{dx^2} +\omega^2-V_{\rm eff}(r)\right)R_{\ell m}(r,\omega) = 0\ ,
\end{equation}
with the effective potential
\begin{equation}\label{Veff}
V_{\rm eff}(r)=f(r)\left[\frac{\ell(\ell+1)}{r^2}+
\frac{f'(r)}{r} + \mu_{\rm eff}^2 \right]
\ ,
\end{equation}
where
\begin{eqnarray}
dx=\frac{dr}{f(r)}, \qquad \mu_{\rm eff}^2 =\mu^2-m\, qB\ .
\end{eqnarray}
From the form of the effective potential one can realize  that in
addition to the squared mass of the scalar field, $\mu^2$, there
is another term $-mqB$. This is due to the well-known Zeeman
effect, which is the shift of energy of charged particle, $q$, in
the magnetic field, $B$, due to the interaction of magnetic and
scalar fields through the charge with an azimuthal momentum $m$.
Let us analyze the symmetry features of the effective potential
through effective mass. Here all the parameters $m$, $q$, and $B$
can be both positive and negative. Therefore, there are the
following physically reasonable combinations of signatures of
these parameters in that effective potential via effective mass
remains invariant:
\begin{itemize}
\item $q\rightarrow-q$, \quad $B\rightarrow-B$;
\item $m\rightarrow-m$, \quad $B\rightarrow-B$;
\item $m\rightarrow-m$, \quad $q\rightarrow-q$.
\end{itemize}
\begin{figure}[ht]
\centering\includegraphics[width=0.43\textwidth]{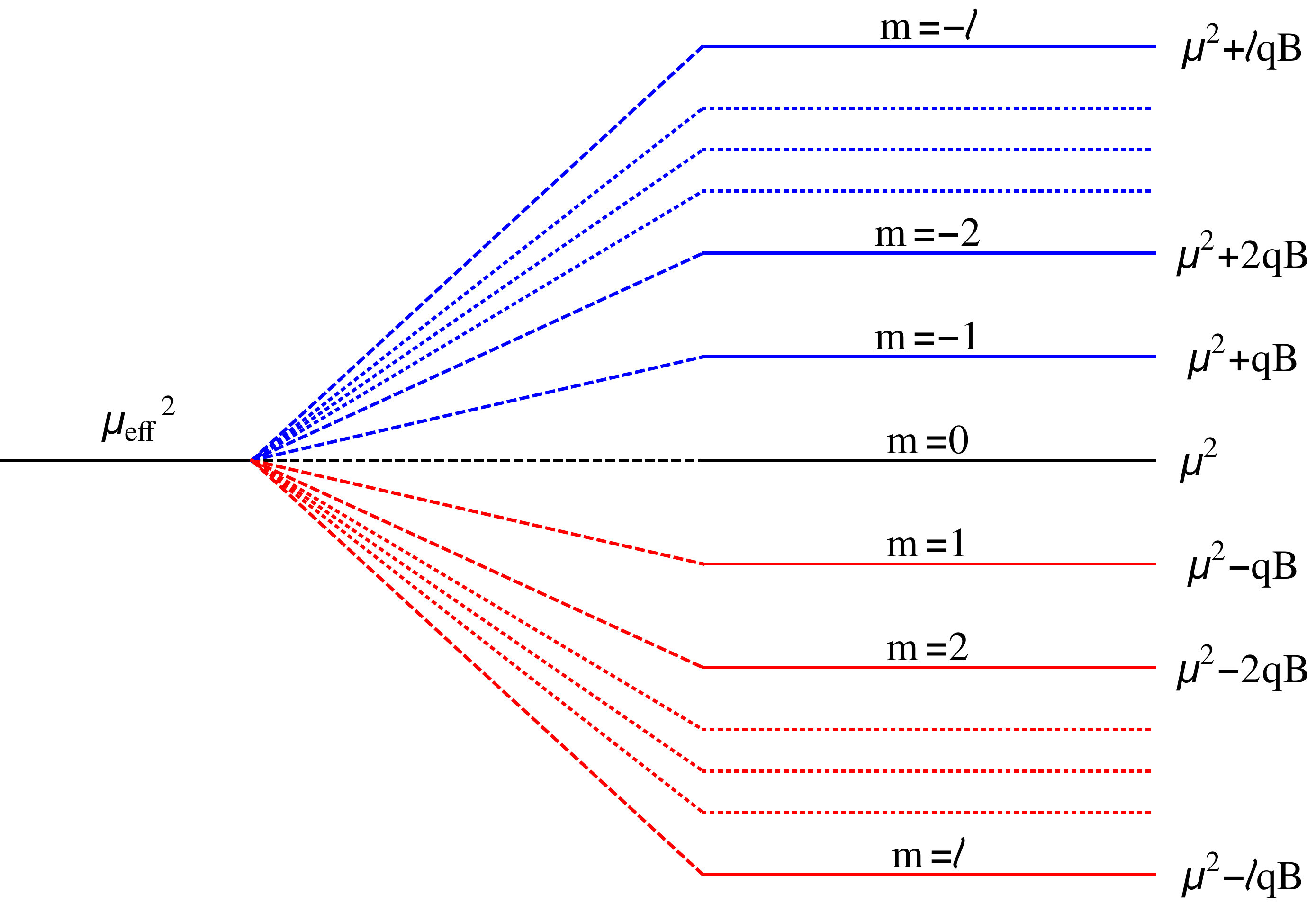}
\caption{\label{effmass} Splitting of the effective mass in
the external magnetic field.}
\end{figure}
As we have mentioned before that the azimuthal number $m$ accepts
$2\ell+1$ values in the region $-\ell\leq m\leq \ell$. Taking this
into account one can see that the effective mass also splits into
$2\ell+1$ parts as schematically shown in Fig.~\ref{effmass} and
consequently, possible $2\ell+1$ numbers of the effective
potentials appear.

\begin{figure}[ht]
\centering\includegraphics[width=0.45\textwidth]{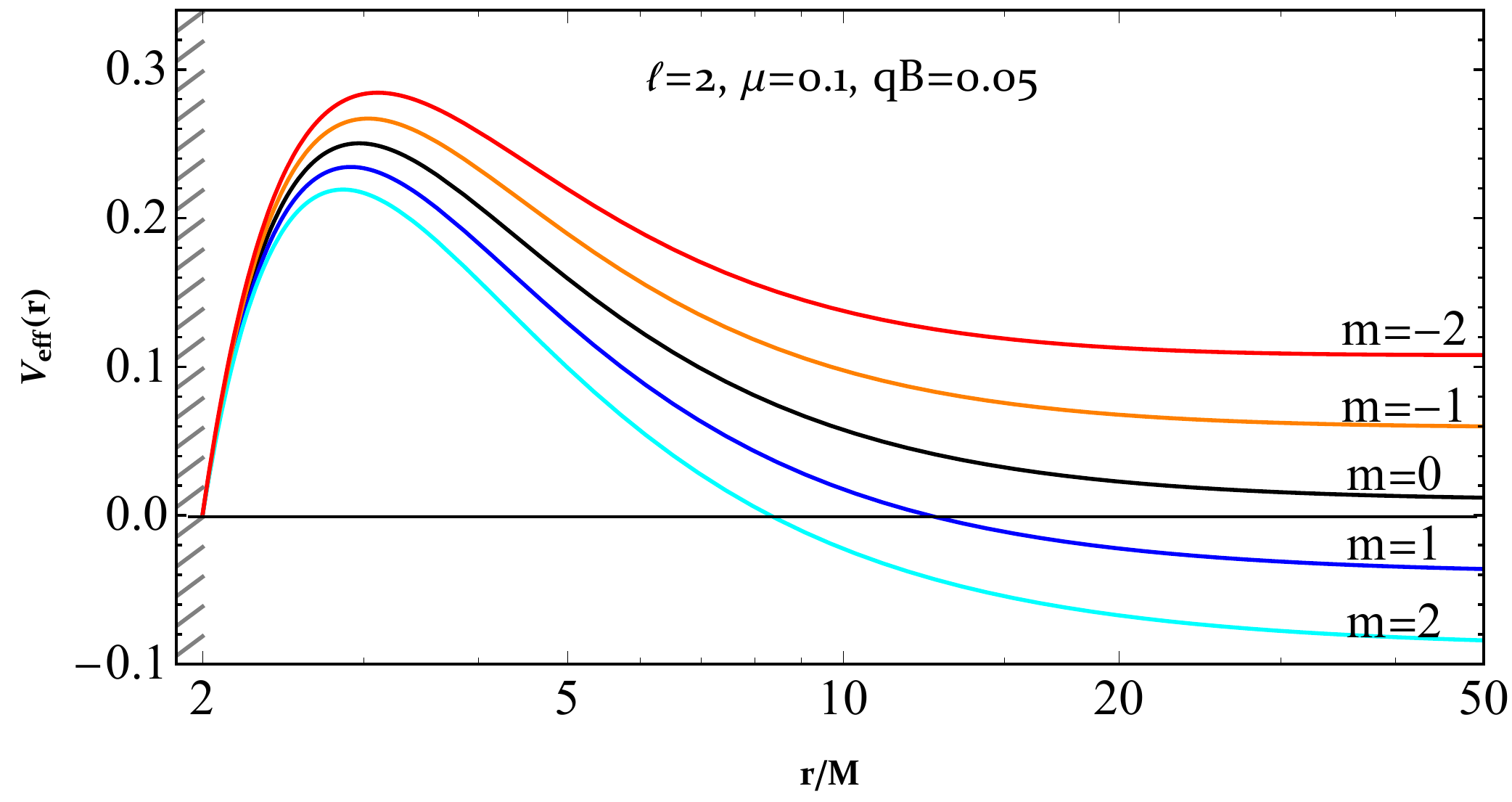}
\caption{\label{Potential} Radial profiles of the all possible
effective potentials for the multipole number $\ell=2$.}
\end{figure}

In similar way, we can analyze the effective potential in the
presence of the external magnetic field. From the expression
(\ref{Veff}) we can see that for the given multipole number
$\ell$, we get $2\ell+1$ different shapes of the effective
potential. For simplicity, here we set the multipole number
$\ell=2$ which means that $-2\leq m\leq 2$. In
Fig.~\ref{Potential} the radial dependence of the effective
potential is illustrated. The black curve corresponds to uncharged
scalar field ($q=0$), and for the charged scalar field ($q\neq0$)
this curve is split into five different ones as shown in
Fig.~\ref{Potential}. Moreover, one should note that the massive
scalar field which satisfies the relation $\mu^2=m qB$ has the
same dynamics as massless one. Note that for some positive values
of azimuthal number, $m$, an instability may exist depending on
the value of mass of the scalar field, $\mu$, further away from
the black hole.

Last term of the effective potential in expression (\ref{Veff})
can be considered as induced mass due to the interaction of the
scalar and magnetic fields and it plays very important role in
calculation of quasinormal modes of the black hole which will be
considered in next section~\ref{Result}. Since, as we have
mentioned before that $m$ can also accept the negative values
($-\ell<m<\ell$), there are the following scenarios related with
interaction of the scalar and magnetic fields which cause either
an increase or decrease in the value of the effective mass:
\begin{itemize}
\item $mqB > 0$\quad {\rm at}\quad $m>0$, \quad $qB>0$\quad or\quad $m<0$,\quad $qB<0$;
\item $mqB < 0$\quad at\quad $m>0$,\quad $qB<0$\quad or\quad $m<0$,\quad $qB>0$.
\end{itemize}

In the papers~\cite{Konoplya08a,Konoplya08b,Wu15} the quasinormal
modes  of uncharged scalar field in the Ernst (strongly magnetized
black hole) spacetime were studied and it was shown that in the
$G^{1/2}c^{-2}B r\ll 1$ approximation, the scalar field
propagating on the Ernst background is equivalent to the massive
scalar perturbation propagating on the Schwarzschild background
with effective mass $\mu_{\rm eff}^2=\mu^2+4B^2m^2$ which is
always positive. Here additional term $\sim4B^2m^2$ does not
characterize interaction between the scalar and the
electromagnetic fields but arises directly due the spacetime
geometry of the magnetized black hole. However, such approximation
is not accurate enough in the strong magnetic and gravitational
fields when $G^{1/2}c^{-2}Br\sim1$~\cite{Brito14}.

\begin{figure}[ht]
\centering\includegraphics[width=0.45\textwidth]{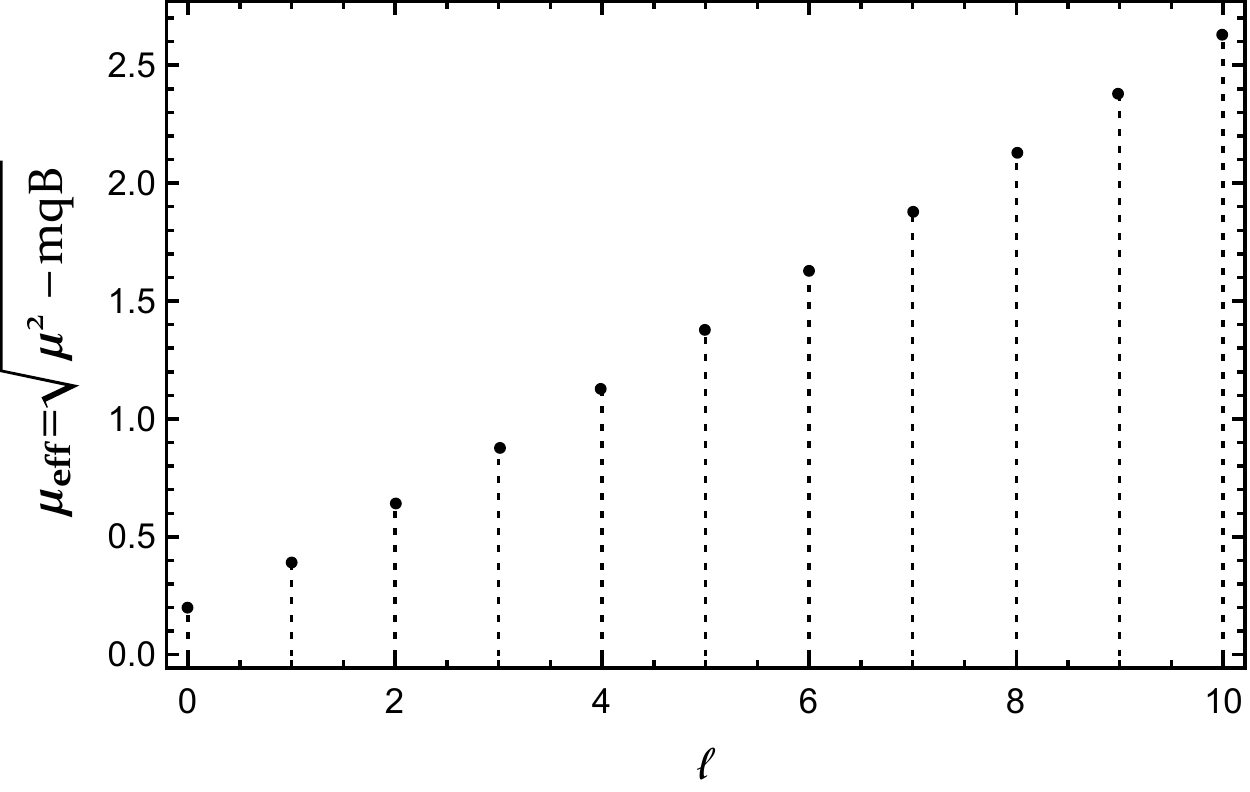}
\caption{\label{mass-limit} Possible values of the effective mass
of  the scalar field that give the limit of existence of
quasinormal modes in the Schwarzschild black hole. Where black
spots correspond to the threshold values at that the quasinormal
modes are arbitrarily long lived. Dashed region represents allowed
values of the effective mass to exist quasinormal modes.}
\end{figure}

One can see from Fig.~\ref{Potential} that at large distances
$r\to \infty$ the effective potential tends to square of the
effective mass as
\begin{eqnarray}
V_{\rm eff}(r\rightarrow\infty)\rightarrow\mu_{\rm
eff}^2\equiv\mu^2-mqB\ .
\end{eqnarray}
It means that depending on the values of mass and charge of the
scalar field,  azimuthal number, and strength of the magnetic
field, value of the squared effective mass can be negative also.
In this case, one can observe an instability. Moreover, if the
value of the effective mass is positive, there is some threshold
value of the effective mass after that the effective potential
loses its barrier-like form and quasinormal modes disappear. At
this threshold value of the effective mass, there are arbitrarily
long lived quasinormal modes, so called quasi-resonance
modes~\cite{Ohashi:CQG:2004,Toshmatov:PRD:2016,Toshmatov:EPJP:2017}.
Since finding of this threshold value analytically is impossible,
we present some values of the threshold effective mass for
different values of the multipole number in Fig.~\ref{mass-limit}.

\section{Numerical results\label{Result}}

In this section we briefly present results of the numerical
calculations of the quasinormal modes of the Schwarzschild black
hole immersed in the external asymptotically  uniform magnetic
field. In order to solve the Schr$\ddot{o}$dinger-like wave
equation (\ref{WaveEq}), we set the following boundary conditions:
asymptotic behaviour of the wave is purely incoming at the event
horizon and purely outgoing at the spatial infinity as
\begin{eqnarray}\label{boundary-conds}
R(r) =
\begin{cases}
e^{-i\omega\, x} \quad
\textrm{at}\quad x\rightarrow-\infty
\, , \ ( r\rightarrow2M)\ ,\\
e^{i\chi\, x} \quad
\textrm{at} \quad x\rightarrow\infty
\, , \ ( r\rightarrow\infty) \ ,
\end{cases}
\end{eqnarray}
where $\chi=\sqrt{\omega^2-\mu_{{\rm eff}}^2}$.

\subsection{WKB method}

For numerical calculations of quasinormal frequencies we use  the
WKB method that was applied for the first time for the
calculations of quasinormal modes of black holes by Schutz and
Will~\cite{Schutz85} and after that the method was extended to
higher orders by several
authors~\cite{Iyer87a,Iyer87b,Konoplya03}. The sixth order WKB
method for solving the Schrdinger-like wave equation governing the
quasinormal modes of black holes implies the relation
\begin{equation}
\frac{i(\omega^2-V(r_0))}{\sqrt{-2V''(r_0)}} + \sum_{j=2}^6 \Lambda_j = n+\frac{1}{2},
\end{equation}
where $r_0$ is the extreme of the potential that is corresponding
to its maximum, a prime ``$'$'' denotes the derivative with
respect to tortoise coordinate $x$, $n$ is the label the overtone,
$\Lambda_j$ are the correction terms that can be found
in~\cite{Schutz85,Iyer87a,Iyer87b,Konoplya03}.

By using the sixth order WKB method, we obtain the frequencies of
quasinormal modes of electrically charged scalar perturbations in
the field of magnetized Schwarzschild black hole. In
Tab.~\ref{Tab} the real and imaginary part of the frequencies of
quasinormal modes for different values of the azimuthal $m$ and
the multipole $\ell$ numbers and different values of the parameter
$qB$ are produced. From the Tab.~\ref{Tab} one can easily see that
spectrum of the quasinormal modes splits into $2\ell+1$ parts due
to the magnetic field in analog to the Zeeman effect. Note that
here we have set the overtone number as $n=0$.
\begin{table*}
\caption{The fundamental ($n=0$) quasinormal frequencies
($M\omega$) of massive,  charged scalar field in the magnetized
Schwarzschild black hole calculated via the sixth order WKB
method. \label{Tab}}
\begin{ruledtabular}
\begin{tabular}{cccccccc}
$\ell$ & $m$ & $\mu=0, qB=0$ & $\mu =0.1, qB=0$ & $\mu = 0.1, qB = 0.05$ & $\mu = 0.1, qB = 0.1$ &$\mu = 0.1, qB = 0.15$ \\ \hline
 0  & 0 & 0.1104 - 0.1008i & 0.1122 - 0.0927i & 0.1122 - 0.0927i & 0.1122 - 0.0927i & 0.1122 - 0.0927i \\
\hline
   & 1 &                  &                  & 0.2752 - 0.1082i & 0.2535 - 0.1206i & 0.2322 - 0.1323i \\
 1 & 0 & 0.2929 - 0.0977i & 0.2974 - 0.0951i & 0.2974 - 0.0951i & 0.2974 - 0.0951i & 0.2974 - 0.0951i \\
   & -1 &                 &                  & 0.3200 - 0.0809i & 0.3429 - 0.0654i & 0.3543 - 0.0441i \\
\hline
   & 2 &                  &                  & 0.4555 - 0.1063i & 0.4249 - 0.1165i & 0.3948 - 0.1263i \\
   & 1 &                  &                  & 0.4711 - 0.1011i & 0.4555 - 0.1063i & 0.4401 - 0.1114i \\
2  & 0 & 0.4836 - 0.0967i & 0.4868 - 0.0957i & 0.4868 - 0.0957i & 0.4868 - 0.0957i & 0.4868 - 0.0957i \\
   & -1&                  &                  & 0.5027 - 0.0902i & 0.5188 - 0.0845i & 0.5351 - 0.0787i \\
   & -2 &                 &                  & 0.5188 - 0.0845i & 0.5516 - 0.0726i & 0.5852 - 0.0598i \\
\hline
   & 3 &                  &                  & 0.6426 - 0.1044i & 0.6081 - 0.1125i & 0.5740 - 0.1205i \\
   & 2 &                  &                  & 0.6542 - 0.1016i & 0.6310 - 0.1071i & 0.6081 - 0.1125i \\
   & 1 &                  &                  & 0.6659 - 0.0988i & 0.6542 - 0.1016i & 0.6426 - 0.1043i \\
3  & 0 & 0.6754 - 0.0965i & 0.6777 - 0.0959i & 0.6777 - 0.0959i & 0.6777 - 0.0959i & 0.6777 - 0.0959i \\
   & -1&                  &                  & 0.6896 - 0.0931i & 0.7015 - 0.0901i & 0.7135 - 0.0872i \\
   & -2 &                 &                  & 0.7015 - 0.0901i & 0.7256 - 0.0842i & 0.7499 - 0.0781i \\
   & -3 &                 &                  & 0.7135 - 0.0872i & 0.7499 - 0.0781i & 0.7872 - 0.0685i \\
\end{tabular}
\end{ruledtabular}
\end{table*}
One can see from Tab.~\ref{Tab} that an increase in the value of
effective mass  increases the frequency of the real oscillations
and decreases damping rate of the scalar field.

\subsection{Leaver's method}

In order to check and increase accuracy of the quasinormal
frequencies,  we adopt alternate method, i.e., the Leaver's method
which is continued fraction method (also known as Leaver's method)
developed in 1985 by Leaver~\cite{Leaver}. In this method one
chooses a solution of Eq. (\ref{WaveEq}) to be in the following
form:
\begin{eqnarray}\label{wave-function}
R(r)=r^{i(\omega^2+\chi^2)/2\chi}(r-1)^{-i\omega}e^{i\chi r}
\sum_{n=0}^{\infty}a_n\left(\frac{r-1}{r}\right)^n\ .
\end{eqnarray}
Here Leaver's unit $2M=1$ is adopted. The coefficients
$a_n\in\{a_0=1,a_1,a_2,...\}$ satisfy the following three-term
recurrence relation:
\begin{eqnarray}
&&\alpha_0 a_1+\beta_0 a_0=0\ ,\\
&&\alpha_na_{n+1}+\beta_na_n+\gamma_na_{n-1}=0\ . \nonumber
\end{eqnarray}
The recurrence coefficient $\alpha_n$, $\beta_n$  and $\gamma_n$
are given by
\begin{eqnarray}
&&\alpha_n=n^2+(c_0+1)n+c_0\ ,\nonumber\\
&&\beta_n=-2n^2+(c_1+2)n+c_3\ ,\\
&&\gamma_n=n^2+(c_2-3)n+c_4-c_2+2\ ,\nonumber
\end{eqnarray}
with
\begin{eqnarray}
&&c_0=1-i2\omega\ ,\nonumber\\
&&c_1=-4+i(4\omega+3\chi)+\frac{i\omega^2}{\chi},\nonumber\\
&&c_2=3-i(2\omega+\chi)-\frac{i\omega^2}{\chi},\nonumber\\
&&c_3=\left(1+\frac{\omega}{\chi}\right)\left[ (\omega+\chi)^2 +\frac{i(\omega+3\chi)}{2}\right]-\ell(\ell+1)-1\ ,\nonumber\\
&&c_4=-\left[i+\frac{(\omega+\chi)^2}{2\chi}\right]^2\ . \nonumber
\end{eqnarray}
For the given multipole number $\ell$ by using the boundary
conditions~(\ref{boundary-conds}) for the
series~(\ref{wave-function}) the frequency $\omega$ is found by
solving the following equation:
\begin{eqnarray}\label{leaver}
\beta_n-&&\frac{\alpha_{n-1}\gamma_n}{\beta_{n-1}-
 \frac{\alpha_{n-2}\gamma_{n-1}}{\beta_{n-2}-\alpha_{n-3}\gamma_{n-2}/...}}\nonumber\\
&&=\frac{\alpha_{n}\gamma_{n+1}}{\beta_{n+1}-
\frac{\alpha_{n+1}\gamma_{n+2}}{\beta_{n+2}-\alpha_{n+2}\gamma_{n+3}/...}}\ .
\end{eqnarray}
This equation is complicated and cannot be solved analytically.
Therefore, we present only numerical results in Tab.~\ref{Tab-II}.
\begin{table*}
\caption{The fundamental quasinormal frequencies of massive,
charged scalar field for the magnetized Schwarzschild black hole
calculated via the Leaver's method with $\mu=0.1$. \label{Tab-II}}
\begin{ruledtabular}
\begin{tabular}{ccccccccc}
$\ell$ & $m$ & $ qB = 0.05$ & $qB = 0.05$ & $qB = 0.1$ & $qB=0.1$  & $qB=0.15$ & $qB=0.15$ \\
 & & (stable) & (unstable) & (stable) & (unstable) & (stable) & (unstable) \\
\hline
 0  & 0 & 0.1110 - 0.1028i & - & 0.1110 - 0.1028i & - & 0.1110 - 0.1028i & - \\ \hline
   & 1 & 0.2885 - 0.1002i & 0.2000i & 0.2829 - 0.1036i & 0.3000i & 0.2774 - 0.1068i & 0.3742i \\
 1 & 0 & 0.2940 - 0.0970i & - & 0.2940 - 0.0970i & - & 0.2940 - 0.0970i & - \\
   & -1 & 0.2997 - 0.0936i & - & 0.3053 - 0.0901i & - & 0.3109 - 0.0866i & -\\
\hline
   & 2 & 0.4766 - 0.0992i & 0.3000i & 0.4687 - 0.1018i & 0.4359i & 0.4609 - 0.1045i & 0.5386i \\
   & 1 & 0.4805 - 0.0978i & 0.2000i & 0.4766 - 0.0992i & 0.3000i & 0.4726 - 0.1005i & 0.3742i \\
2  & 0 & 0.4844 - 0.0965i & - & 0.4844 - 0.0965i & - & 0.4844 - 0.0965i & - \\
   & -1 & 0.4884 - 0.0951i & - & 0.4923 - 0.0938i & - & 0.4963 - 0.0924i & - \\
   & -2 & 0.4923 - 0.0938i & - & 0.5003 - 0.0910i & - & 0.5083 - 0.0882i & - \\
\hline
   & 3 & 0.6671 - 0.0985i & 0.3742i & 0.6583 - 0.1006i & 0.5385i & 0.6496 - 0.1027i & 0.6633i \\
   & 2 & 0.6701 - 0.0978i & 0.3000i & 0.6642 - 0.0992i & 0.4359i & 0.6583 - 0.1006i & 0.5386i \\
   & 1 & 0.6730 - 0.0971i & 0.2000i & 0.6701 - 0.0978i & 0.3000i & 0.6671 - 0.0985i & 0.3742i \\
3  & 0 & 0.6760 - 0.0964i & - & 0.6760 - 0.0964i & - & 0.6760 - 0.0964i & -  \\
   & -1 & 0.6789 - 0.0956i & - & 0.6819 - 0.0949i & - & 0.6848 - 0.0942i & -  \\
   & -2 & 0.6819 - 0.0949i & - & 0.6878 - 0.0935i & - & 0.6937 - 0.0920i & -  \\
   & -3 & 0.6848 - 0.0942i & - & 0.6937 - 0.0920i & - & 0.7027 - 0.0898i & -  \\
\end{tabular}
\end{ruledtabular}
\end{table*}
By comparing numerical results given in Tabs.~\ref{Tab}
and~\ref{Tab-II}  that are derived by the WKB and Leaver's
methods, respectively, one can easily notice the significant
difference in the quasinormal frequencies for negative values of
the squared effective mass, $\mu_{\rm eff}^2<0$. In the results
evaluated by the WKB method given in Tab.~\ref{Tab} imaginary
parts of the frequencies are always negative which are an
indication of stability of the spacetime against such
perturbations, while in Tab.~\ref{Tab-II} calculated by the
Leaver's method, in addition to the stable modes which are almost
the same as the one calculated by the WKB method in
Tab.~\ref{Tab}, unstable modes with positive imaginary frequency
appear. Existence of both stable and unstable modes for the
negative squared effective mass with the same multipole is
strange. Therefore, one needs to exclude one of them. In order to
determine whether the unstable modes are physically relevant or
not, in the next subsection we study the temporal evolution of the
perturbations.

\subsection{Instability}

One can see from the effective potential depicted in
Fig.~\ref{Potential}  that for the negative values of the
effective mass, $\mu_{\rm eff}^2<0$, the effective potential is
negative in the region $r\in(r_n,\infty)$. Although, an increase
in the value of multipole number increases the height of the
potential, it does not change depth of the negative part of the
effective potential, whereas it pushes away this negative region,
i.e., it increases $r_n$. The depth of negativity of the effective
potential is only characterized by  effective mass. Despite
negativity of the effective potential indicates an instability, it
does not guarantee it~\cite{Konoplya:PRD:2008}. Therefore, to
determine if there is the instability in this region, we
investigate the temporal evolution of the scalar perturbation by
following the time-domain integration method proposed by Gundlach
et al.~\cite{Gundlach:PRD:1994}. The time-domain integration
method involves the retarded $du\equiv dt-dx$ and advanced
$dv\equiv dt+dx$  light-cone variables. Then, taking into account
that the scalar field is harmonically time dependent, the wave
equation (\ref{WaveEq}) is written in the form
\begin{equation}
- 4 \frac{\partial^2 \Phi }{ \partial u\partial v}=V_{\rm eff}(r(u, v)) \Phi\ .
\end{equation}
This equation is solved numerically. To solve this  equation we
use the following discretization scheme:
\begin{equation}
\Phi_N=(\Phi_W+\Phi_E)\frac{16-\Delta^2 V_S}{16+\Delta^2 V_S}-\Phi_S+O(\Delta^4)\ ,
\end{equation}
where the indices $N$, $W$, $E$, and $S$ refer to grid-points
$N\equiv(u,v)$, $W\equiv(u-\Delta,v)$, $E\equiv(u,v-\Delta)$, and
$S\equiv(u-\Delta,v-\Delta)$. $\Delta$ is a constant separating
neighboring points of the grid (for details, please see
\cite{Rezzolla:CQG:2007,TSSA:PRD:2018}). The initial data is
specified on the two null surfaces $u=u_0$ and $v=v_0$. For
calculations we assume that the initial perturbation is Gaussian
pulse centered around the point $v_c$ with width $\sigma$ as
\begin{eqnarray}
\Phi(u=u_0,v)=A\exp\left(-\frac{(v-v_c)^2}{\sigma^2}\right).
\end{eqnarray}
\begin{figure}[ht]
\centering\includegraphics[width=0.48\textwidth]{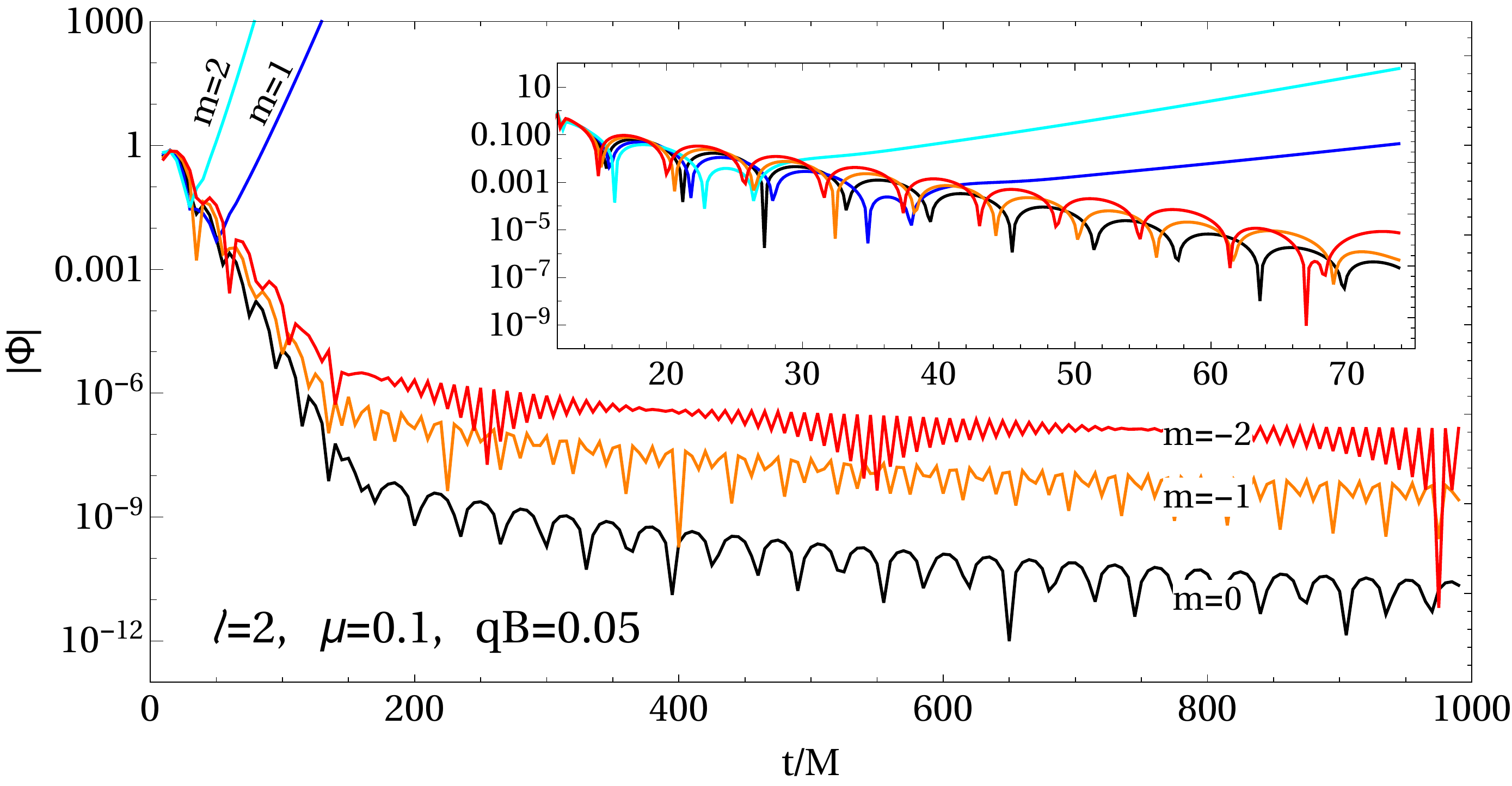}
\caption{\label{TimeEvaluation} Temporal evolution of the charged,
massive scalar field around Schwarzschild black hole immersed in
the external uniform magnetic field.}
\end{figure}
In Fig.~\ref{TimeEvaluation}, evolution of the charged, massive
scalar  perturbation with time are illustrated for the mode
$\ell=2$ and all possible five values of $m$. Due to the external
magnetic field quasinormal modes of the scalar field around black
hole live long in comparison with the quasinormal modes without
magnetic field. It is also shown that the effects of the magnetic
field is more profound in the final stage of the quasinormal
modes. One can see from Fig.~\ref{TimeEvaluation} that instability
occurs for the negative values of the squared effective mass as it
was stated in~\cite{Chen:PLB:2009,Kokkotas:PRD:2011}, in
particular, for the positive value of $qB$ parameter ($qB>0$) and
for positive value of the azimuthal number $m>0$, only if
$\mu_{\rm eff}^2<0$. From the symmetric property of the effective
potential for the values of $m$ and $qB$ one can find instability
in the case of $m<0$ and $qB<0$ too. Thus, the negativity of
effective potential leads the instability.

Moreover, as we have indicated at the end of the previous
subsection,  here we briefly discuss the reason of the
contradiction on the quasinormal modes derived by the WKB and
Leaver's methods which are given in Tabs.~\ref{Tab} and
\ref{Tab-II}, respectively. As it has been perfectly explained in
\cite{Kokkotas:PRD:2011}, in the case of the single peak effective
potential, the WKB method is built on the basis of matching of
three wave solutions at the turning points~\footnote{Turning
points are the solutions of equation $\omega^2-V_{\rm eff}(r)=0$.}
and therefore, it is more accurate only in the region close to
maximum of the effective potential. Therefore, the WKB method can
be applied only near to the maximum of the potential and in our
case it can be used for calculation of the stable modes when the
effective potential does not have negative region. On the
contrary, fortunately, the Leaver's method does not have this
problem and can be applied for the whole region. Consequently, the
Leaver's method provides not only stable but also unstable modes
for the given multipole moment $(l,m)$ as shown in Tab.
\ref{Tab-II}. However, the stable modes produced by the Leaver's
method correspond to the ones of the WKB method which does not
take into account the negative region of the effective potential.
Therefore, for the cases of negative squared effective mass,
stable modes should be considered as just mathematical solution
having no physical meaning. With increase of the negativity of the
effective mass, instability also increases. Interestingly, this
instability does not depend on multipole number of the scalar
field, being characterized by only effective mass of the charged,
massive scalar field.

\section{Conclusion}\label{Conclusion}

In this paper we have investigated the dynamics of charged and
massive scalar field in the vicinity of static spherically
symmetric black hole immersed into the external  asymptotically
uniform magnetic field. Assuming the scalar field as spherical
symmetric and harmonically time dependent as well as using the
general relativistic form of the Klein-Gordon equation we have
obtained stationary Schr\"{o}dinger-like equation with effective
potential which possesses the squared effective mass, $\mu_{\rm
eff}^2=\mu^2-mqB$ produced by the Zeeman effect that represents a
shift of energy of charge of the scalar field $q$ in the magnetic
field $B$ due to the interaction of magnetic field with an
azimuthal momentum $m$. By investigating the effective potential
we have found early indication of the instability that occurs for
negative values of the squared effective mass. Since azimuthal
number satisfies the condition $|m|\leq\ell$, for given $\ell$ the
squared effective mass of the scalar field $\mu_{\rm
eff}^2=\mu^2-mqB$ and consequently, the quasinormal modes are
splitted into $2\ell+1$ parts due to the magnetic Zeeman effect.
Moreover, the threshold value of the effective mass of the scalar
field from which the effective potential loses its barrier-like
form and quasinormal modes disappear has been found. Analysis of
temporal evolution of the scalar field has shown that for the
negative values of the squared effective mass, $\mu_{\rm
eff}^2<0$, there is instability. However, we have shown that the
WKB method is not able to show this instability as the negative
part of the effective potential is located further away from the
peak of the potential. Therefore, in order to show the instability
we have used the Leaver's method. It has been shown that due to
the external magnetic field and only in the case when the
effective mass is positive, quasinormal modes are becoming longer
lived and have larger oscillation frequencies for the positive
charged scalar field until the effective mass reaches the
threshold value.

\section*{Acknowledgements}
B.T. is grateful to Roman Konoplya for discussions. The research
is supported in part by Czech Science Foundation GA\v{C}R project
No.~19-03950S, Nazarbayev University Faculty Development
Competitive Research Grant No.~090118FD5348, Uzbekistan Ministry
for Innovational Development Grants No.~VA-FA-F-2-008,
No.~MRB-AN-2019-29 and No.~YFA-Ftech-2018-8, the Abdus Salam
International Centre for Theoretical Physics through Grant
No.~OEA-NT-01 and Erasmus+ exchange grant between Silesian
University in Opava and National University of Uzbekistan. B.T.
acknowledges the internal student grant of Silesian University in
Opava No. SGS/12/2019. B.A. acknowledges Fudan Fellowship towards
his stay at the Fudan University, Shanghai, China.

\bibliographystyle{apsrev4-1}
\bibliography{Ref}

\end{document}